\documentclass[aps,pra,showpacs,superscriptaddress,preprint]{revtex4}
\usepackage{amsmath}
\usepackage{epsf}
\usepackage{epsfig}
\usepackage{bm}

\catcode`ð=\active
\defð{\u{g}}
\catcode`Ð=\active
\defÐ{\u{G}}
\catcode`Ý=\active
\defÝ{\. I}
\catcode`ö=\active
\defö{\"{o}}
\catcode`Ö=\active
\defÖ{\"O}
\catcode`ü=\active
\defü{\"{u}}
\catcode`Ü=\active
\defÜ{\"{U}}
\catcode`Þ=\active
\defÞ{\c{S}}
\catcode`þ=\active
\defþ{\c{s}}
\catcode`ý=\active
\defý{{\i}}
\catcode`ç=\active
\defç{\d{c}}
\catcode`Ç=\active
\defÇ{\d{C}}

\begin{document}
\title{Effective-mass Klein-Gordon Equation for non-PT/non-Hermitian Generalized
Morse Potential}
\author{\small Altu\={g} Arda}
\email[E-mail: ]{arda@hacettepe.edu.tr}\affiliation{Department of
Physics Education, Hacettepe University, 06800, Ankara,Turkey}
\author{Ramazan Sever}
\email[E-mail: ]{sever@metu.edu.tr}\affiliation{Department of
Physics, Middle East Technical University, 06531, Ankara,Turkey}

\date{\today}

\begin{abstract}

The one-dimensional effective-mass Klein-Gordon equation for the
real, and non-\textrm{PT}-symmetric/non-Hermitian generalized
Morse potential is solved by taking a series expansion for the
wave function. The energy eigenvalues, and the corresponding
eigenfunctions are obtained. They are also calculated for the constant mass case.\\
Keywords: position dependent mass, Klein-Gordon Equation,
generalized Morse potential, \textrm{PT}-symmetry, energy
eigenvalues, eigenfunctions.
\end{abstract}

\pacs{03.65.Fd, 03.65.Ge}

\maketitle

\newpage

\section{Introduction}

In the past few decades, there has been an increased interest to
find the exact solutions of the non-relativistic and relativistic
equations. The solutions of the Schrödinger equation has been
studied by using different methods based on perturbative and
non-perturbative approaches [1-5]. The Klein-Gordon (KG) and Dirac
equations have been also studied for different type of potentials
such as Aharonov-Bohm (AB) potential [6], the AB plus Dirac
monopole potential [7, 11], generalized Hulthen, harmonic, and
linear potentials, generalized asymmetrical Hartmann potentials,
for a uniform magnetic field, pseudoharmonic oscillator, and
exponential-type potentials [12-20].

Recently, there has been considerable work on the
\textrm{PT}-symmetric quantum mechanics. Following the initially
studies of Bender et al. [21], the \textrm{PT}-symmetric theory
has been successfully studied by many authors because of their
importance, and also for using in different research areas, such
as quantum field theory, and nuclear physics [22, 23]. In the
present study, we take into account the form of the generalized
Morse potential having also non-\textrm{PT} symmetric property to
achieve the solutions within the position-dependent mass
formalism, and study the effects of the mass varying with
spatially coordinate on the solutions of the case of constant
mass. It could be interested to solve the relativistic KG equation
in this point of view.

The solutions of the above wave equations are taken in general for
constant or time-dependent masses [24, 25]. There are also many
examples of physically important systems, for which the mass
depends on coordinates. So far, solving the above equations for
the case of spatially dependent mass has been studied by many
authors [26-34] for different types of the mass distributions such
as an exponentially, and quadratic mass functions [27],
inverse-quadratic dependence of mass [35], trigonometric
mass-distributions [36], and mass functions of the forms
m(r)=r$^{\alpha}$, especially for three-dimensional problems [36,
37]. We will use also a mass distribution having an exponential
form, and study in the half-plane to achieve a physically
acceptable (finite) mass function, which has a decreasingly
behavior in this range.

In the present work, we study the bound state solutions of the KG
equation with real and non-\textrm{PT} symmetric/non-Hermitian
forms of generalized Morse potential in the case of a
coordinate-dependence mass distribution. In order to find the
energy spectra, and the corresponding wave functions we prefer to
use an infinite series for the wave function. This approach is a
powerful technique to solve second order differential equations
especially for the potential forms having two and/or more terms
[38, 39]. We obtain also the energy spectrum, and the
corresponding eigenfunctions in the case of constant mass for two
potential types.

The organization of the work as follows. In section II, we obtain
the exact bound-state energy spectra for real, and non-\textrm{PT}
symmetric/non-Hermitian cases of generalized Morse potential, and
the corresponding eigenfunctions by taking an infinite series for
the wave function in the case of a convenient mass distribution.
We summarize the numerical results in one table, and two figures
to see the effects of the mass depending on coordinate. For this
aim we do the numerical analysis for three different diatomic
molecules such as \textrm{H}$_2$, \textrm{LiH}, and \textrm{HCl}.
We give also the results for the case of constant mass, and
present our conclusions in Section III.

\section{Bound State Solutions}
\subsection{Real Generalized Morse Case}

The one-dimensional KG equation is written in terms of scalar and
vector potential as following [40]

\begin{eqnarray}
\frac{\textrm{d}^2\psi(x)}{\textrm{d}x^2}\,+\,\frac{1}{\hbar^2c^2}\,\left[\left[E-V_{v}(x)\right]^2-
\left[mc^2+V_{s}(x)\right]^2\right]\psi(x)=0\,,
\end{eqnarray}
where \textrm{m} is the mass of the particle, \textrm{E} denotes
the energy, and \textrm{V}$_s(x)$ and \textrm{V}$_v(x)$ are the
scalar and vector parts of the potential, respectively, and
\textrm{c} is the velocity of light. In the absence of the vector
potential, the KG equation can be written as

\begin{eqnarray}
\frac{\textrm{d}^2\psi(x)}{\textrm{d}x^2}\,+\,\frac{2m}{\hbar^2}\,
\left(\varepsilon-\frac{V_{s}^2(x)}{2mc^2}-V_{s}(x)\right)\psi(x)=0\,,
\end{eqnarray}
which is a Schrödinger-like equation with the energy eigenvalue
$\varepsilon=\frac{E^2}{2mc^2}-\frac{mc^2}{2}$.

We consider the scalar potential V$_s(x)$ in Eq. (2) as
generalized Morse potential, which can be used to describe the
vibrations of a two-atomic molecule, as the following

\begin{eqnarray}
V_s(x)=V_1 \textrm{e}^{-2\beta x}-V_2\textrm{e}^{-\beta
x}\,,\,\,(0 \leq x \leq +\infty)\,.
\end{eqnarray}
where V$_1$, and V$_2$ are two real parameters, and
$x=(r-r_0)/r_0$. The parameter $\beta$ is ar$_0$, here, a is the
potential width, and r$_0$ is the equilibrium distance.

Here, we prefer to use the following mass-distribution

\begin{eqnarray}
m(x)=m_0+m_1\textrm{e}^{-\beta x}\,,
\end{eqnarray}
where m$_0$, and m$_1$ are real parameters. This distribution is
finite at infinity, and enables us to analyze the results in the
case of constant mass.

Substituting Eqs. (3) and (4) into Eq. (2), than we have

\begin{eqnarray}
\frac{\textrm{d}^2\psi(x)}{\textrm{d}x^2}\,&+&\Big\{\,\frac{1}{\hbar^2c^2}\,[E^2-m_0^2c^4]
+\,\frac{2}{\hbar^2c^2}\,[V_2m_0c^2-m_0m_1c^4]\textrm{e}^{-\beta
x}\nonumber
\\&-&\,\frac{1}{\hbar^2c^2}\,[V_{2}^2+2V_1 m_0 c^2-2V_2 m_1c^2+m_{1}^2 c^4]\textrm{e}^{-2\beta
x}\nonumber
\\&+&\,\frac{2}{\hbar^2c^2}\,[V_1V_2-V_1m_1c^2]\textrm{e}^{-3\beta
x}-\,\frac{V_1^2}{\hbar^2c^2}\,\textrm{e}^{-4\beta
x}\Big\}\psi(x)=0\,,
\end{eqnarray}

By using the transformations on coordinate and wave function

\begin{eqnarray}
z=\textrm{e}^{-\beta
x}\,,\,\,\,\,\,\,\,\psi(z)=\,\frac{1}{\sqrt{z}}\,\phi(z)\,,
\end{eqnarray}
and with the help of the following parameters

\begin{eqnarray}
A_1&=&Q^2(E^2-m_0^2c^4)\,+\,\frac{1}{4}\,,\nonumber \\
A_2&=&2Q^2(V_2m_0c^2-m_0m_1c^4)\,,\nonumber \\
A_3&=&Q^2(V_{2}^2+2V_1 m_0 c^2-2V_2 m_1c^2+m_{1}^2
c^4)\,,\nonumber \\
A_4&=&2Q^2(V_1V_2-V_1m_1c^2)\,,\nonumber \\
A_5&=&Q^2V_1^2\,.
\end{eqnarray}
we get

\begin{eqnarray}
\frac{\textrm{d}^2\phi(z)}{dz^2}\,+\,\left(-A_3+\,\frac{A_1}{z^2}\,+\,\frac{A_2}{z}\,
-A_4 z-A_5z^2\right)\phi(z)=0\,.
\end{eqnarray}
where Q$^2=1/\hbar^2c^2\beta^2$ in the above expressions $(0 \leq
z \leq 1)$. It would be interested the results obtained from the
last equation for a special case A$_{2}$=A$_{4}=0$, which gives

\begin{eqnarray}
\frac{\textrm{d}^2\phi(z)}{dz^2}\,+\,\left(\frac{A_1}{z^2}-A_5z^2\right)\phi(z)=A_{3}\phi(z)\,,
\end{eqnarray}

This equation has a similar form with the ones given by Eq. (7) in
Ref. [17] for A$_{5} \rightarrow \alpha^2$, A$_{1} \rightarrow
-\ell'(\ell'+1)$, and A$_{3} \rightarrow -\lambda$ ($\alpha,
\ell'$, and $\lambda$ are the parameters used in Ref. [17]). Eq.
(9) corresponds to the equation of the harmonic oscillator with
centrifugal potential barrier, so we could give the solutions as

\begin{eqnarray}
\varepsilon_{n}=\mp\Big\{m^2_0 c^4+\frac{4c^2}{\beta^2\hbar^2}(m_1
c^2+m_0)^2\big[2n+1+\sqrt{\frac{1}{4}-V^2_1Q^2\,}\big]^{-2}-\frac{1}{4Q^2}\Big\}^{1/2}\,.
\end{eqnarray}

We turn to the solution of Eq. (8), and write the wave function as
[38, 39]

\begin{eqnarray}
\phi(z)=\textrm{e}^{pz+(1/2)qz^2}\sum_{n=0}^{\infty} a_n
z^{2n+L+1/2}\,.
\end{eqnarray}

Substituting Eq. (11) into Eq. (8), and equating of the
coefficients to zero, we get the following identities among
coefficients, and the expressions for p, and q

\begin{eqnarray}
X_n a_n+Y_{n+1}a_{n+1}+Z_{n+2}a_{n+2}=0\,,
\end{eqnarray}
where

\begin{eqnarray}
X_n&=&2q(2n+L+1)-A_3\,,\\
Y_n&=&A_2+p\,(4n+2L+1)\,,\\
Z_n&=&4n(n+L)+2L^2\,.
\end{eqnarray}
and

\begin{eqnarray}
q^2&=&A_5\,\\
2pq&=&A_4\,.
\end{eqnarray}
where the new parameter L is defined as L$^2 \rightarrow
Q^2(E^2-m_0^2c^4)$, and we choose the parameters as
p$=Q(V_2-m_1c^2)$, and q$=-QV_1$ to obtain a physically solution.

On the other hand, X$_n$, Y$_n$ and Z$_n$ must satisfy the
following condition to determine the coefficients in the system of
equations given in Eqs. (13)-(15)

\begin{eqnarray}
\textrm{det}\left| \begin{array}{cccccc} Y_0 & Z_1 & \ldots &
\ldots
&\ldots & 0 \\ X_0 & Y_1 & Z_2 & \ldots & \ldots & 0 \\ X_1 & Y_2 & Z_3 & \ldots & \ldots & 0 \\
\vdots & \vdots & \vdots & \ddots & \vdots & \vdots \\
0 & 0 & 0 &0 & X_{n-1} & Y_n \end{array} \right|=0\,,
\end{eqnarray}

In the case of a$_n \neq0$, but a$_{n+1}=$a$_{n+2}= \ldots=0$ in
Eq. (12), we impose X$_n=0$. This leads to the following algebraic
equation

\begin{eqnarray}
2q(2n+L+1)=A_3\,.
\end{eqnarray}

Substituting the values of  the parameters defined in Eq. (7) into
Eq. (19), and by using p, and q obtained from Eqs. (16)-(17), we
have the energy eigenvalues of the generalized Morse potential as

\begin{eqnarray}
E_n=\pm\left\{m_0^2c^4\,+\,\frac{1}{Q^2}\left[2n+1+Q\tilde{V_1}\right]^2
+\,\frac{m_1c^2\tilde{V_2}}{Q}\,\left[2(2n+1)+2Q\tilde{V_1}+Qm_1c^2\tilde{V_2}\right]\right\}^{1/2}\,,\nonumber\\
\end{eqnarray}
where

\begin{eqnarray}
\tilde{V_1}=\,\frac{V_2^2}{2V_1}\,+m_0c^2\,,\,\,\,\,\,\,\tilde{V_2}=\,-\frac{V_2}{V_1}\,+\,\frac{m_1c^2}{2V_1}\,,
\end{eqnarray}

It is seen that the energy levels of the particle and
antiparticles are split around zero. On the other hand, the energy
spectrum of the generalized Morse potential for the position-
dependent mass is real. The part of the energy eigenvalues coming
from the coordinate dependence of the mass is dependent on the
quantum number n, and also on the potential parameters V$_1$, and
V$_2$.

The corresponding wave functions can be written as the following

\begin{eqnarray}
\phi_n(z)=(a_0z^{1/2}+a_1z^{5/2}+
\ldots)z^{L}\,\textrm{exp}\left[Q(V_2-m_1c^2)z-\,\frac{QV_1}{2}\,z^2\right]\,.
\end{eqnarray}
where a$_{\textrm{i}} (\textrm{i}=0,1 \ldots)$ are the
coefficients of the series.

In order to obtain the energy spectrum of the real generalized
Morse potential for the case of constant mass, we have to set
m$_1=0$ in Eq. (20), than we get

\begin{eqnarray}
E^{m_1=0}_n=\pm\left(2n+1+Q\tilde{V_1}\right)\,\sqrt{\,\frac{1}{Q^2}\,+
\left(\,\frac{m_0c^2}{2n+1+Q\tilde{V_1}}\right)^2}\,,
\end{eqnarray}

According to Eq. (23), the energy levels of the particle and
antiparticles are symmetric about the zero in the limit of the
constant mass. On the other hand, the energy spectrum of the real
generalized Morse potential is purely real as in the first case.

We summarize the numerical results in Table I for the H$_2$,
\textrm{LiH}, and \textrm{HCl} molecules in the case of the
constant mass m$_0$ obtained from Eq. (23). We give the bound
state energies of the particles (+E$_n$), and antiparticles
(-E$_n$) for different n-values in Table I. Further, we plot the
dependence of the ground states of the above molecules on m$_1$
for H$_2$, and \textrm{LiH}, \textrm{HCl} molecules in Fig.1, and
Fig.2, respectively. We do the numerical analysis in the range
$10^{-6} < \frac{1}{M} < 10^{-4}$, where M=m$_0/$m$_1$, and (p)
denotes the particle case, and (a) denotes the antiparticle one in
figures.

The corresponding eigenfunctions in the case of the constant mass
can be obtained from Eq. (22) by setting m$_1=0$.

\subsection{Non-\textrm{PT} Symmetric and non-Hermitian Generalized Morse Case}
Now we consider the complex case of the generalized Morse
potential, where the potential parameters are written as

\begin{eqnarray}
V_1&=&
\upsilon_1^2-\upsilon_2^2+2\textrm{i}\upsilon_1\upsilon_2\,,\nonumber
\\V_2&=&\upsilon_1+\textrm{i}\upsilon_2+2\upsilon_3(\upsilon_1+\textrm{i}\upsilon_2)\,.
\end{eqnarray}
where $\upsilon_{\textrm{i}}'s (\textrm{i}=1, 2, 3)$ are the real
parameters, and $\beta=1$. This complex form of the generalized
Morse potential is called non-\textrm{PT} symmetric/non-Hermitian
potential form of the potential [41, 42].

Substituting Eq. (24) into Eq. (21), than we have from Eq. (20)
for the energy spectra of the non-\textrm{PT}
symmetric/non-Hermitian generalized Morse potential the following

\begin{eqnarray}
E_n=\pm\left\{m_0^2c^4\,+\,\frac{1}{Q'^2}\left[2n+1+Q'\tilde{V'_1}\right]^2
+\,\frac{m_1c^2\tilde{V'_2}}{Q'}\,\left[2(2n+1)+2Q'\tilde{V'_1}+Q'm_1c^2\tilde{V'_2}\right]\right\}^{1/2}\,,\nonumber\\
\end{eqnarray}
where

\begin{eqnarray}
\tilde{V'_1}=\,\frac{2(\upsilon_3+1)^2}{2}\,+m_0c^2\,,\,\,\,
\tilde{V'_2}=\,\frac{2\upsilon_3+1}{\upsilon_1+\textrm{i}\upsilon_2}\,+\,
\frac{m_1c^2}{2(\upsilon_1+\textrm{i}\upsilon_2)^2}\,,\,\,\,Q'=\,\frac{1}{\hbar^2c^2}\,.
\end{eqnarray}

The corresponding eigenfunctions can be written as

\begin{eqnarray}
\phi_n(z)=(a_0z^{L+1/2}+a_1z^{L+5/2}+
\ldots)\,\textrm{exp}\left[Q(V_2-m_1c^2)z-\,\frac{QV_1}{2}\,z^2\right]\,.
\end{eqnarray}

The energy spectrum for the case of constant mass is given as

\begin{eqnarray}
E^{m_1=0}_n=\pm\left(2n+1+Q'\tilde{V'_1}\right)\,\sqrt{\,\frac{1}{Q'^2}\,+
\left(\,\frac{m_0c^2}{2n+1+Q'\tilde{V_1}}\right)^2}\,,
\end{eqnarray}

It is clearly seen that the energy eigenvalues of the
non-\textrm{PT} symmetric/non-Hermitian generalized Morse
potential has real and imaginary parts, which is coming from the
coordinate dependence of the mass, but in the constant mass limit
the energy spectrum is real. On the other hand, the energy levels
of the non-\textrm{PT} symmetric/non-Hermitian generalized Morse
potential is symmetric about zero as in the real potential case.
The eigenfunctions for the case of constant mass are obtained by
setting m$_1=0$ in Eq. (27).

\section{Conclusion}

We have obtained the solutions of the KG equation for the real and
non-\textrm{PT} symmetric/non-Hermitian generalized Morse
potential with position dependent mass. The energy spectra and the
corresponding wave functions have been obtained by using an
infinite series for the wave function. We have found that the real
generalized Morse potential has a real energy eigenvalues in the
case of the position-dependent mass. We have summarized first the
energy eigenvalues of three diatomic molecules for different
values of $n$, and than plot two different graphs where the energy
eigenvalues varying with mass to see the effects of the mass
depending on spatially coordinate. We have also studied the
non-\textrm{PT} symmetric/non-Hermitian case of the potential, and
pointed out that the energy eigenvalues are imaginary in this
case. We have found that the energy levels are purely real in the
case of the constant mass as given in Eqs. (23), and (28) for both
form of the generalized Morse potential.

\section{Acknowledgments}

This research was partially supported by the Scientific and
Technical Research Council of Turkey.

\newpage

\begin{table}
\caption{\label{tab:special}The dependence of bound states on n in
$MeV$ for H$_2$ (D=38267.76 cm$^{-1}$, a=1,9426$ \AA^{-1}$,
r$_0=0,7416 \AA$, and m$_0=0,50391$amu), LiH (D=20287 cm$^{-1}$,
a=1,1280$ \AA^{-1}$, r$_0=1,5956 \AA$, and m$_0=0,8801221$ amu),
and \textrm{HCl} molecule (D=37255 cm$^{-1}$, a=1,8677$ \AA^{-1}$,
r$_0=1,2746 \AA$, and m$_0=0,9801045$ amu) [43] with $\ell=0$.}
\begin{ruledtabular}
\begin{tabular}{cccc}
$n$ & $\pm$ E$_n$(H$_2)$ & $\pm$ E$_n$(LiH) & $\pm$ E$_n$(HCl)\\
0 & 663,819 & 1159,420 & 1291,130\\
2 & 663,827 & 1159,430 & 1291,140\\
4 & 663,835 & 1159,440 & 1291,150\\
10 & 663,859 & 1159,470 & 1291,190\\
20 & 663,899 & 1159,520 & 1291,260\\
30 & 663,939 & 1159,570 & 1291,330\\
40 & 663,979 & 1159,620 & 1291,390\\
50 & 664,020 & 1159,670 & 1291,460\\
\end{tabular}
\end{ruledtabular}
\end{table}

\newpage

\begin{figure}[htbp]
\centering
\includegraphics[height=4.5in, width=7in,
angle=0]{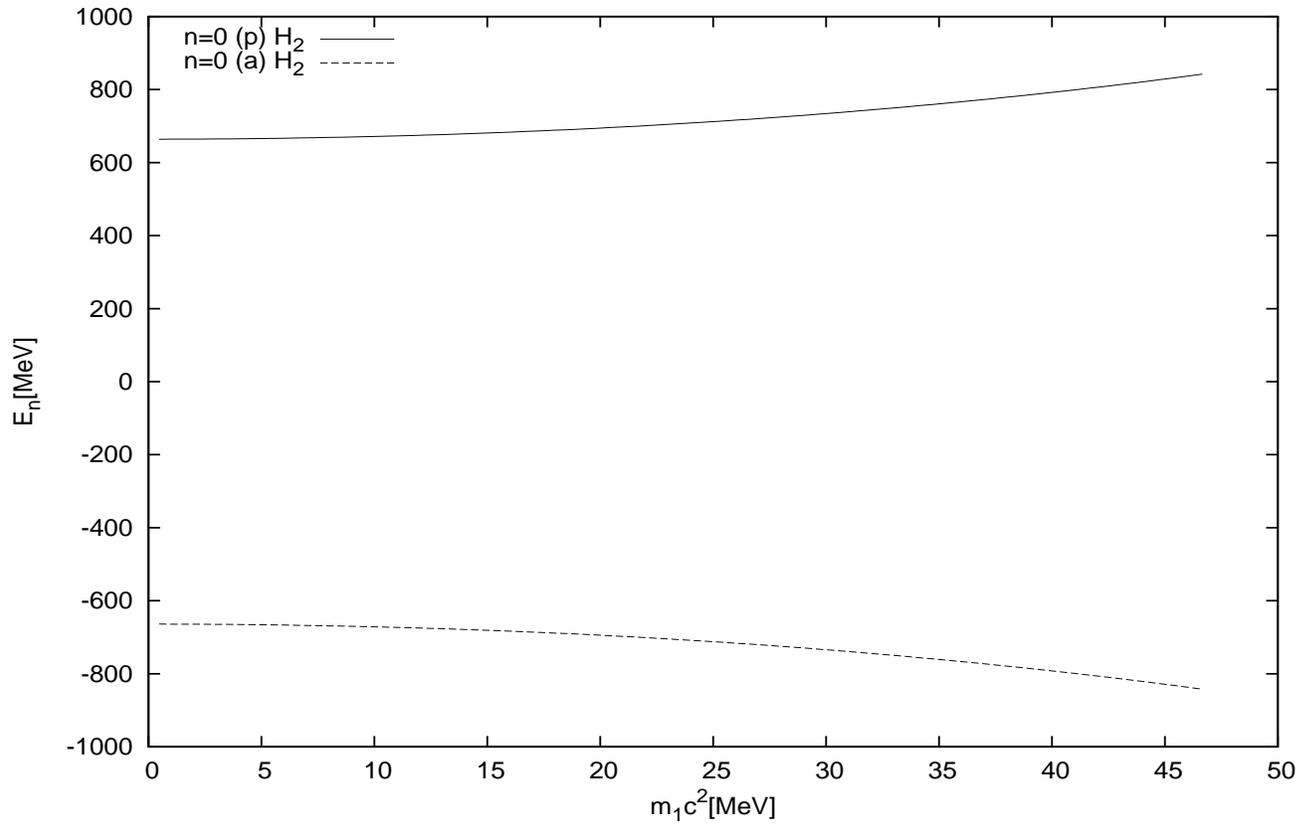} \caption{The variation of ground states of the
H$_2$ molecule with m$_1$.}
\end{figure}

\begin{figure}[htbp]
\centering
\includegraphics[height=4.5in, width=7in, angle=0]{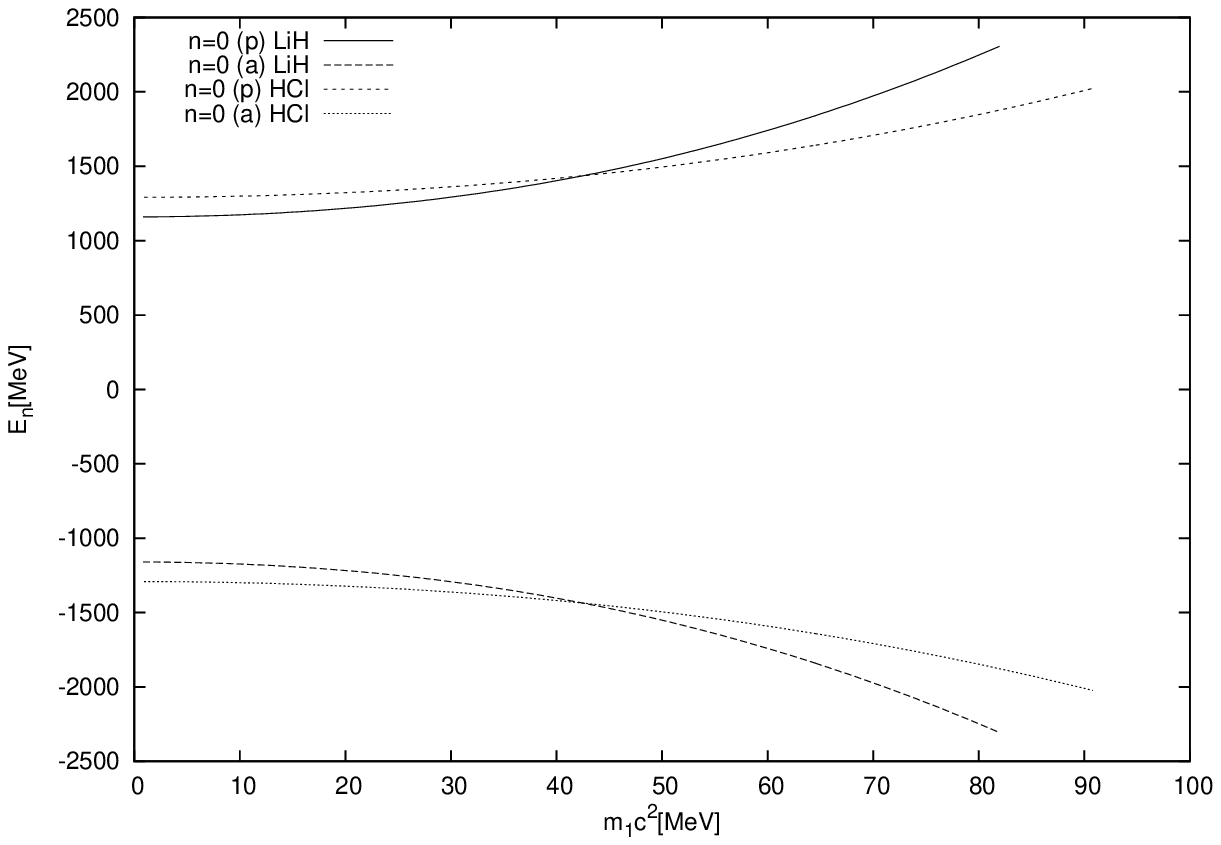}
\caption{The variation of ground states of the \textrm{LiH}, and
\textrm{HCl} molecules with m$_1$.}
\end{figure}


\newpage



\begin{thebibliography}{99}



\bibitem{ref1} J.~P.~Killingbeck, A.~Grosjean, and G.~Jolicard,
J. Phys. A: Math. Gen. {\bf 34}, 8309 (2001).



\bibitem{ref2} A.~Arda, Turk. J. Phys. {\bf 28}, 223 (2004).







\bibitem{ref3} J.~Chen, L.~C.~Kweck, and C.~H.~Oh, Phys. Rev. A {\bf 67}, 012101
(2003).



\bibitem{ref4} S.~N.~Biswas et al., J. Math. Phys. {\bf 14}, 1190 (1973).







\bibitem{ref5} A.~Khare, and U.~P.~Sukhatme, J. Phys. A {\bf 26}, L901-L904
(1993) [arXiv: hep-th/9212147].







\bibitem{ref6} C.~R.~Hagen, Phys. Rev. D {\bf 48}, 5935 (1993),
[arXiv:~hep-th/9308009].



\bibitem{ref7} V.~M.~Villalba, [arXiv:~hep-th/9409102].



\bibitem{ref8} A.~S.de Castro, Int. J. Mod. Phys. A {\bf 22}, 2609 (2007) [arXiv:~hep-th/0511010].



\bibitem{ref9} A.~S.de Castro, Phys. Lett. A {\bf 338}, 81 (2005),
[arXiv:~hep-th/0502201].



\bibitem{ref10} A.~S.de Castro, Phys. Lett. A {\bf 346}, 71 (2005),
[arXiv:~hep-th/0507218].



\bibitem{ref11} A.~S.de Castro, Phys. Lett. A {\bf 342}, 53 (2005) [arXiv:~hep-th/0409216].



\bibitem{ref12} W.~C.~Qiang, R.-S.~Zhou, and Y.~Gao, Phys. Lett. A {\bf 371}, 201 (2007).



\bibitem{ref13} N.~Saad, [arXiv:~math-ph/0709.4014].



\bibitem{ref14} R.~Giachetti, and E.~Sorace, Phys. Rev. Lett. {\bf 101}, 190401 (2008) [arXiv:~hep-th/0706.0127].



\bibitem{ref15} K.~Bhattacharya, [arXiv:~hep-th/0705.4275].



\bibitem{ref16} A.~S.~Dutra, and M.~B.~Hott, [arXiv:~quant-ph/0705.3447].



\bibitem{ref17} C.~Gang, C.~Zi-Dong, and L.~Zhi-Mei, Chi. Phys. {\bf 13}, 279
(2004).



\bibitem{ref18} C.~Gang, Phys. Lett. A {\bf 339}, 300 (2005).



\bibitem{ref19} C.~Gang, Acta. Phys. Sin. {\bf 50}, 1651 (2001).



\bibitem{ref20} L.~Z.~Yi, Y.~F.~Diao, J.~Y.~Liu, and C.~S.~Jia, Phys. Lett. A {\bf 333}, 212 (2004).



\bibitem{ref21} C.~M~.~Bender, S.~Boettcher, and P.~N.~Meisinger, J. Math. Phys. {\bf 40}, 2201 (1999).



\bibitem{ref22} S.~M.~Ikhdair, and R.~Sever, Ann. Phys. {\bf 16}, 218 (2007).



\bibitem{ref23} C.~S.~Jia, P.~Q.~Wang, J.~Y.~Liu, and S.~He,
Int. J. Theor. Phys. {\bf 47}, 2513 (2008), and references
therein.



\bibitem{ref24} A.~Mostafazadeh, J. Phys. A {\bf 31}, 6495 (1998).



\bibitem{ref25} A.~Mostafazadeh, Phys. Rev. A {\bf 55}, 4084 (1997).



\bibitem{ref26} A.~D.~Alhaidari, Phys. Lett. A {\bf 322}, 72 (2004).



\bibitem{ref27} A.S.~Dutra, and C.~A.~S.~Almeida, Phys. Lett. A {\bf 275}, 25 (2000).



\bibitem{ref28} B.~Gonul, B.~Gonul, D.~Tutcu, and O.~Ozer, Mod. Phys. Lett. A {\bf 17}, 2057 (2002).



\bibitem{ref29} B.~Gonul, O.~Ozer, B.~Gonul, and F.~Uzgun, Mod. Phys. Lett. A
{\bf 17}, 2453 (2002).



\bibitem{ref30} B.~Gonul, and M.~Kocak, Chin. Phys. Lett. {\bf 20}, 2742
(2005).



\bibitem{ref31} B.~Gonul, and M.~Kocak, [arXiv:~quant-ph/0512035].



\bibitem{ref32} C.~Tezcan, and R.~Sever, J. Math. Chem. {\bf 42}, 387 (2007) [arXiv:~quant-ph/0604041].



\bibitem{ref33} I.~O.~Vakarchuk, J. Phys. A: Math. Gen. {\bf 38}, 4727 (2005).



\bibitem{ref34} C.~Gang, and Z.~Chen, Phys. Lett. A {\bf 331}, 312 (2004).



\bibitem{ref35} L.~Jiang, L.-Z.~Yi, and C.-S.~Jua, Phys. Lett. A {\bf 345}, 249 (2005).



\bibitem{ref36} A.~D.~Alhaidari, Phys. Rev. A {\bf 66}, 042116 (2002).



\bibitem{ref37} G.~X.~Ju, Y.~Xiang, and Z.~Z.~Ren, [arXiv: quant-ph/0601005].



\bibitem{ref38} S.~H.~Dong, Z.~Ma, and G.~Espozito, Found. Phys. Lett. {\bf 12},
465 (1999).



\bibitem{ref39} S.~H.~Dong, Int. J. Theor. Phys. {\bf 39}, 1119 (2000); {\bf
40}, 569 (2001).



\bibitem{ref40} A.~S.~de Castro, Phys. Lett. A {\bf 338}, 81 (2005).



\bibitem{ref41} Z.~Ahmed, Phys. Lett. A {\bf 282}, 343 (2001).



\bibitem{ref42} G.~Bagchi, and C.~Quesne, Phys. Lett. A {\bf 273}, 285 (2000).



\bibitem{ref43} E.~D.~Filho, and R.~M.~Ricotta, Phys. Lett. A {\bf 269}, 269 (2000) [arXiv: hep-th/9910254].
\end{thebibliography}
\end{document}